\documentclass[aps,prd,preprint,nofootinbib,showpacs,preprintnumbers,showkeys]{revtex4}

\usepackage{epsfig}
\usepackage[dvipdfm]{hyperref} 

\def\e{\,{\rm e}\,}
\def\d{{\rm d}}
\def\i{{\rm i}}

\def\C{C}
\def\L{R}
\def\D{{\cal D}}
\def\Tau{{\cal T}}
\def\x{r}

\newcommand{\rf}[1]{(\ref{#1})}
\newcommand{\eq}[1]{Eq.~(\ref{#1})}

\def\be{\begin{equation}}
\def\ee{\end{equation}}
\def\bea{\begin{eqnarray}}
\def\eea{\end{eqnarray}}
\def\LA{\left\langle}
\def\RA{\right\rangle}

\newcommand{\ie}{{i.e.}\ }

\def\la{\lesssim}
\def\ga{\gtrsim}

\begin{document}

\preprint{ITEP--TH--11/11}

\title{An interplay between static potential and\\  Reggeon trajectory 
for QCD string} 
\thanks{Based on a talk at the meeting of the Royal Danish Cycling Society.}

\author{Yuri Makeenko}

\affiliation{Institute of Theoretical and Experimental Physics\\
B.~Cheremushkinskaya 25, 117218 Moscow, Russia}
\email{makeenko@itep.ru} 

\date{March 11, 2011}

\begin{abstract}
I consider two cases where QCD string is described by an effective theory 
of long strings:
the static potential and meson scattering amplitudes in the Regge regime.
I show how they can be solved in the mean-field approximation, 
justified by the large number of space-time dimensions, and argue that
it turns out to be exact.
I compare contributions from QCD string and perturbative QCD
and discuss experimental consequences for the scattering amplitudes.

\end{abstract}



\maketitle

\section{Introduction}

QCD string is formed at the distances larger than the confinement scale and 
is described by an effective string theory, which has a well-defined 
semiclassical expansion around a long-string ground state.
The leading order, known as the L\"uscher term~\cite{LSW80},
is universal and possesses a number of remarkable properties, 
owing to the fact that the transverse size of the long string is also large.
The number of transverse degrees of freedom of the effective string resides 
in an overall coefficient of the L\"uscher term, 
which is the only input parameter that equals
$d\!-\!2$ for a bosonic string in $d$ space-time dimensions. 
A long-standing idea is to extract physical consequences directly 
from this well-established picture of QCD string.

I consider in this Letter two cases where QCD string is outstretched,
so that the effective string theory is applicable:
a rectangular Wilson loop, determining the static potential, and
meson scattering amplitudes in the Regge kinematical regime,
determining the Reggeon trajectory.
I emphasize similarities between these two cases and describe how
they can be solved in the mean-field approximation, 
which is justified by large $d$, 
but turns out to be exact at any $d$.
I take into account contributions to the scattering amplitudes
from both QCD string and perturbative QCD
and discuss some experimental consequences.

\section{Static potential}

A linear static potential $V(R)$ between heavy quarks in QCD corresponds to
the area-law behavior of a $R\times T$ rectangular Wilson loop 
\be
W\left(R\times T \right) = \e^{-T V(R)}
\label{potential}
\ee 
for $T\gg R$
and results in quark confinement. This implies that the gluon field forms
a flux tube or a string for $R$ larger than the confinement scale.

In string theory a linear potential is associated with a classical string,
while a quantization of an outstretched string
with the Dirichlet boundary condition in $d$ dimensions
yields~\cite{Alv81,Arv83}
\be
V(R)=\frac{1}{2\pi \alpha'}
 \sqrt{\L^2- \pi^2 \frac{(d-2)}{6}\alpha'},
\label{Vstring}
\ee
where $1/2\pi\alpha'$ is the string tension. 
Equation~\rf{Vstring} was first obtained~\cite{Alv81}
in the large-$d$ limit, where only transverse degrees of freedom are
essential, and then derived~\cite{Arv83} by the Virasoro quantization 
of the string with the Dirichlet boundary condition. 
What was most important
with the formula in \eq{Vstring} was a recognition of the fact that
it to be exact for any $d$  order by order in $1/\L$, 
since the Lorentz anomaly vanishes at large $R$
so the quantization is consistent~\cite{Ole85}.
The root in \eq{Vstring} becomes imaginary for small distances, which
is related to tachyonic instability, but there is no tachyon for long strings.  

For a covariant quantization the Lorentz anomaly is translated into
the Virasoro anomaly.
It is now possible to describe QCD string in $d<26$ by an effective string 
theory with a nonpolynomial action~\cite{PS91}
\begin{equation} 
S_{\rm eff}=\frac{1}{\pi\alpha'}
\int\d^2 z \,\partial X \cdot \bar \partial X + 
\frac{(d-26)}{24\pi}  \int \d^2 z \, 
\frac{\partial^2 X \cdot  \bar \partial^2 X}{\partial X \cdot \bar \partial X}
+\ldots, 
\label{effaction}
\end{equation} 
where the conformal anomaly is expressed (modulo total derivatives and 
the constraints) via an induced metric 
\be
\e^{\varphi_{\rm ind} }=2 \,\partial X \cdot \bar \partial X ,
\label{indmetric}
\ee
which is not treated
independently as distinct from the Polyakov formulation. 
It was shown~\cite{PS91,Dru04} by analyzing the effective string theory
order by order in $1/R$ that the
Virasoro anomaly vanishes and the spectrum reproduces that~\cite{Arv83} 
of the Nambu--Goto string in $d$ dimensions,
which remarkably agrees with the recent lattice calculations~\cite{Tep09}. 

\section{Momentum-space disk amplitude}

The consideration of scattering amplitudes for QCD string is pretty much
similar to the above analysis of the static potential.
They are given by a momentum-space disk amplitude for (smeared) 
stepwise~\cite{MO10b}
\be
p^\mu(t)=\frac 1\pi\sum_i  p_i^\mu \arctan \frac{(t-t_i)}{\varepsilon_i}
 \;{\rightarrow}
\frac1{2} \sum_i  p_i^\mu \,{\rm sign}\,(t-t_i),
\label{stepx}
\ee
where $-\infty < t < +\infty$ is a parametrization of the momentum-space
loop $p^\mu(t)$ and $\varepsilon_i$ will play the role of a regularization. 
The (smeared) discontinuities of the step function~\rf{stepx} correspond to
momenta of colliding particles (all of them are considered as incoming).
Such a momentum-space loop is closed as a consequence of the momentum
conservation. 
For $d=26$, when the action is quadratic, this momentum-space
amplitude coincides with the usual coordinate-space disk amplitude for 
the polygonal loop
\be
x^\mu (t)= 2\pi \alpha' p^\mu(t),
\label{xp}
\ee 
whose vertices $x^\mu_i$ obey $x^\mu_{i+1}-x^\mu_i=2 \pi \alpha' p_i^\mu$.
It looks like the polygon with light-like edges
in Ref.~\cite{AM07a} for ${\cal N}=4$ super Yang--Mills.

For a $2\to2$ process it is convenient to consider a $u$-channel kinematics,
when Mandelstam's variables $s,t<0$ and $u>0$. Keeping in mind an application to
high-energy scattering, we can set 
$p_i^2=0$ for $-s\ga -t\gg  p_i^2$.
At the classical level, we consider a harmonic extension of the
boundary function~\rf{stepx} into the upper half-plane (UHP) $z=x+\i y$:
\be
X^\mu(x,y)=  2 \alpha'\sum_i p_i^\mu  
\arctan \frac{(x-s_i)}{y},
\label{amostX}
\ee
where $s_i=s(t_i)$ for a certain reparametrizing function $s(t)$
obeying $\d s /\d t\geq 0$.

For the function~\rf{amostX} the quadratic part of the action~\rf{effaction} 
equals 
\be
\frac1{2\pi\alpha'} 
S_{\rm quad} = \alpha' s \ln \x+\alpha' t \ln (1-\x),
\label{Squa}
\ee
where
\be
\x=\frac{s_{43} s_{21}}{s_{42} s_{31}},\qquad s_{ij}=s_i-s_j
\label{defr}
\ee
is the projective-invariant ratio.
The minimal surface is obtained by minimizing \rf{Squa} with respect to
$\x$, which gives
\be
\x_*=\frac{s}{s+t}.
\ee
This is nothing but the well-known saddle point of the Veneziano
amplitude, that reproduces the Regge behavior
\be
A\propto \e^{-S_{\rm min}/2\pi\alpha'}
\stackrel{s\gg t}\to \e^{\alpha' t \ln ( s/t)}.
\label{Reggestick}
\ee

\section{Semiclassical Reggeon intercept}

The Regge behavior~\rf{Reggestick} with a linear trajectory 
$\alpha(t)=\alpha' t$ of zero intercept is associated with a classical string.
For a long string, quantum fluctuations can be taken into
account in a semiclassical approximation resulting in 
the L\"uscher term~\cite{LSW80}, which is well-known for 
 a $R\times T$ rectangle with $ {T}\gg {R }$:
\begin{equation}
W(\C)
\stackrel{{\rm rectangle}} 
\propto \e^{- {RT}/2\pi\alpha'+\pi(d-2)T/24R}.
\end{equation} 
We shall now demonstrate how this shifts the Reggeon intercept.

UHP can be mapped onto a rectangle by 
the Schwarz--Christoffel formula 
\be
\omega(z)= \sqrt{s_{42}s_{31}}\int_{s_2}^z \frac{\d x } 
{\sqrt{(s_4-x)(s_3-x)(x-s_2)(x-s_1)}},
\label{SCmap}
\ee
where 
the normalization factor is introduced for the projective symmetry.
The new variable $\omega$ takes values inside a $\omega_R\times \omega_T$ 
rectangle with 
\be
\omega_R=2 K\left(\sqrt{1-\x}\right)\stackrel{\x\to1}\to
 \ln\frac{16}{1-\x},\qquad
\omega_T=2 K\left(\sqrt{\x}\right)\stackrel{\x\to1}\to \pi,
\label{elliK}
\ee
where $K$ is the complete elliptic integral of the first kind.
This has the meaning  
of a worldsheet parametrization. 

To calculate the L\"uscher term, we decompose
\be
X^\mu(\omega_1,\omega_2)=X^\mu_{\rm cl}
(\omega_1,\omega_2)+Y^\mu_{\rm q}(\omega_1,\omega_2),
\label{deco}
\ee
where $X^\mu_{\rm cl}$ is harmonic with the boundary value~\rf{stepx}, 
so $Y^\mu_{\rm q}$ has the mode expansion
\be
Y^\mu_{\rm q}(\omega_1,\omega_2) = \sum_{m,n} \chi^\mu_{m n}
\sin \frac{\pi m \omega_1}{\omega_R}  \sin \frac{\pi n \omega_2}{\omega_T }.
\ee
Now the L\"uscher term results from the determinant coming from the
path integral over $Y^\mu_{\rm q}$.

It is clear that each set of modes results in the L\"uscher term
\be
\frac{\pi \omega_T}{24 \omega_R} = \frac 1{24} \ln \frac{16 s}t
\label{TRvsst}
\ee
for $\omega_T\gg \omega_R$. 
There are $(d-2)$ such sets, so their contribution
to the intercept of the Regge trajectory is~\cite{Jan01,Mak11}
\be
\alpha(0)=\frac{d-2}{24}.
\label{alpha0}
\ee 

It worth commenting on the way how the (same)
L\"uscher term emerges for the UHP parametrization, 
when the path integral over $Y^\mu_{\rm q}$ does {\em not}\/ depend on 
the boundary contour $p^\mu(t)$. In the critical 
dimension $d=26$ it comes~\cite{MO10a} now from the path integral over
reparametrizations (or the boundary metrics), while for $d\neq 26$ 
the classical part $X^\mu_{\rm cl}$ in the decomposition~\rf{deco} additionally
contributes through the second term on the right-hand side of \eq{effaction},
rather than the quantum part $Y^\mu_{\rm q}$:%
\footnote{One may wonder why the second term in the action~\rf{effaction}
does not contribute to the L\"uscher term for a rectangle, as the
first term does. This can be verified by evaluating the contribution 
of the second term to the determinant,
coming from the quadratic fluctuations of $Y^\mu_{\rm q}$ in
\eq{deco}, whose trace of the log involves $\zeta(-2)=0$ contributed
by the second term.}
\be
\alpha(0)=1+ \frac{d-26}{24}=\frac{d-2}{24},
\ee
reproducing \eq{alpha0}. 
How this happens for plane contours is described in the 
original paper~\cite{LSW80}, where the determinant of the
Laplace operator is expressed via the metric induced by 
the conformal mapping $w(z)$. 
For this reason the calculation of the L\"uscher term for
UHP is pretty much similar
to that of Ref.~\cite{DOP84} for the contribution of the Liouville
field in the Polyakov formulation. This is because the Liouville field 
can be simply substituted to the given order of the semiclassical expansion 
by its value given by the induced metric~\rf{indmetric}.

\section{Mean-field approximation}

As was shown by Alvarez~\cite{Alv81}, a path integral of
the Nambu--Goto string for a $T\times R$ rectangle is calculable in the limit of
the large $d$ by a saddle-point technique.
The saddle-point value of the (mean-field) action is%
\footnote{We keep here and below $(d-2)$ for the physical number of
transverse degrees of freedom.}
\be
\frac{1}{2\pi \alpha'}S_{\rm mf}{}_*=\frac{T}{2\pi \alpha'}
 \sqrt{\L^2- \pi^2 \frac{(d-2)}{6}\alpha'},
\label{S_*}
\ee
reproducing~\eq{Vstring}.
As is already mentioned, the Virasoro quantization of the Dirichlet string
shows~\cite{Arv83} this formula to be exact and the quantization is 
consistent~\cite{Ole85} order by order in $1/R$. 
This apparently means that the mean field approximation,
which is usually justified by large $d$, turns out
to be {\it exact}\/ for any $d>2$.

Equation~\rf{S_*} can be easily rederived in the conformal gauge using the
worldsheet parametrization, when the expansion goes around
the mean-field configuration 
\be
X^1_{\rm cl}=\frac {\omega_1}{\omega_R}R, \qquad
X^2_{\rm cl}=\frac {\omega_2}{\omega_T}T.
\label{Xcl}
\ee
Here, $\omega_R$ and $\omega_T$ change under reparametrizations 
and have to be considered as variational parameters. Only the ratio
$\omega_T/\omega_R$ will be essential in what follows.

The mean-field action then takes the form
\be
\frac{1}{2\pi \alpha'}S_{\rm mf}=\frac{1}{4\pi \alpha'}
\left(R^2\frac{\omega_T}{\omega_R}+ T^2\frac{\omega_R}{\omega_T}\right)
-\frac{\pi(d-2)}{24}\frac{\omega_T}{\omega_R},
\label{Smfconf}
\ee
where we have also accounted for the L\"uscher term.
There are no corrections to this formula as $R^2\sim d\to\infty$.
Minimizing \eq{Smfconf} with respect to $\omega_T/\omega_R$,
we find
\be
\left(\frac{\omega_T}{\omega_R}\right)_*= 
\frac{T}{\sqrt{\L^2- \pi^2\frac{(d-2)}6\alpha'}}.
\ee
The substitution into \eq{Smfconf} now reproduces \eq{S_*}.

The above mean-field calculation can be repeated for the UHP
parametrization, which is commonly used for representing scattering amplitudes
in string theory, by making use of the Schwarz--Christoffel mapping~\rf{SCmap}. 
The ratio $\omega_T/\omega_R$, given by the ratio of elliptic integrals
in \eq{elliK}, is known as the
Gr\"otzsch modulus which is monotonic in $\x$. For this reason the varying 
with respect to $r$ is equivalent to the varying 
with respect to $\omega_T/\omega_R$ and \eq{S_*} is reproduced.

We are now in a position to consider the scattering amplitude, when 
the Mandelstam variables $s$ and $t$ play the role of $T$ and $R$.
We then have
\be
\frac{1}{2\pi \alpha'}S_{\rm mf}=
\alpha' s \ln \x + \alpha' t \ln (1-\x) +\frac{(d-2)}{24}\ln(1-\x),
\label{SquaLu}
\ee
where we have added to \eq{Squa} the associated momentum-space 
L\"uscher term~\cite{Jan01,Mak11}.
There are again no corrections to \eq{SquaLu} as $R^2\sim d\to \infty$.

Minimizing the right-hand side of \eq{SquaLu} with respect to $\x$, we find
\be
\x_*=\frac{s}{s+t+(d-2)/24 \alpha'}
\ee
which results for $s\gg-t$ in the Regge behavior
\be
A\propto \e^{\alpha(t) \ln (s/t)}
\label{ReggeA}
\ee
with the linear trajectory
\be
\alpha(t)=\frac{(d-2)}{24}+\alpha' t .
\label{linear}
\ee
It is obtained for large $d$ but is expected to be exact for any $d$ as
is already pointed out. The arguments in favor of this conjecture are:\\
1) it is true in the semiclassical approximation;\\
2) it reproduces the exact result in $d=26$;\\
3) it agrees with the existence~\cite{Hoo74b} of a massless bound state in 
 $d=2$ large-$N$ QCD for massless quarks.

The quadratic fluctuations around this mean field are stable for
$\alpha(t) <0$, \ie for
\be
-\alpha' t > \frac{d-2}{24}.
\label{stable}
\ee

\section{Mean-field approximation (continued)}

It is instructive to compare 
the form~\rf{ReggeA} of the momentum-space disk amplitude 
with \eq{potential} for the $R\times T$ rectangular Wilson loop.
Then $\ln (s/t)$ is an analog of $T$ and $-\alpha(t)$ is an analog of $V(R)$.
An important difference is however that \eq{Vstring} involves the square
root, while $\alpha(t)$ is linear in $t$. 

As is well known, the potential \rf{Vstring} is ill-defined for
$R<R_c=\pi\sqrt{(d-2)\alpha'/6}$ because of the tachyonic singularity.
What happens at $R=R_c$ can be understood by calculating 
the ratio of the area of the dominant surface at the saddle point to
the minimal area spanned by the rectangle~\cite{Alv81}
\be
\LA \sqrt{\det g} \RA_{\rm mf}=\frac{2\pi (\alpha')^2}{S_{\rm min}} 
\frac{\d}{\d \alpha'} \ln W
=
\frac{1-\lambda}{\sqrt{1-2\lambda}},\qquad
\lambda=\pi^2\frac{(d-2)\alpha'}{12R^2}.
\label{ratioW}
\ee
It diverges when $R\to R_c$ from above, which means that typical
surface becomes very large and the mean-field approximation ceases to
be applicable.

The situation is different for the scattering amplitude, when
the ratio of the area of the dominant surface at the saddle point to
the minimal area spanned by the polygon is
\be
\LA \sqrt{\det g} \RA_{\rm mf}=
\frac{\alpha(t)}{\alpha't}=1+\frac{(d-2)}{24\alpha't}.
\label{ratioA}
\ee
It vanishes rather than diverges when $-\alpha't$ is approaching the
value on the right-hand side of \eq{stable} from above, so the linear
Reggeon trajectory~\rf{linear}
can be analytically continued to smaller values of $-\alpha't$.

\section{Applicaton to QCD}

QCD string is stretched between quarks, when they are moved apart,
and makes sense of an effective string theory. 
Meson scattering amplitudes in large-$N$ QCD 
are expressed through a sum 
over paths of the Wilson loop, which reduces~\cite{MO08} to the above
momentum-space disk amplitude%
\footnote{The string disk amplitude is associated as usual with planar 
QCD diagrams and the $\bar q q$ Reggeon trajectory.}, 
when quarks are massless and/or the
number of colliding particles is very large. Therefore,
the above Reggeon trajectory is of physical relevance for QCD.

The linear trajectory~\rf{linear} cannot extend to very large
values of $-\alpha't$, where perturbative QCD (pQCD) is applicable.
It was shown~\cite{KL83} that the reggeization of the $\bar q q$ 
trajectory in pQCD is due to double logarithms
coming (in the axial gauge) from ladder diagrams
and resulting in the $2\to2$ amplitude
\be
A=\frac{2 I_1 \left(\omega \ln (s/\mu^2) 
\right)}{\omega \ln {(s/\mu^2)} }-1\propto \e^{\omega \ln (s/\mu^2)}
\qquad  \omega=\sqrt{\frac{{g}^2(t) C_F}{2\pi^2}}\approx .5,
\ee
where $I_1$ is the Bessel function and $\mu\sim 1$~GeV is an infrared cutoff 
in pQCD. Strictly speaking, the running of $g^2(t)$ with $t$ 
is beyond the accuracy of the calculation, 
but it is expected from asymptotic freedom and
provides the asymptote $\omega\to0$ as $t\to-\infty$, which follows
from the quark counting rule.

In the sum-over-path representation of meson scattering amplitudes
in QCD the separation of pQCD and QCD string can be performed by
splitting the integral over the proper time into two domains:
$\Tau\la1/\mu$ and $\Tau\ga1/\mu$. The domain $\Tau\la1/\mu$ is
associated with small paths and, correspondingly, pQCD, while  
 domain $\Tau\ga1/\mu$ is
associated with large paths and, correspondingly, QCD string.
The total amplitude is the sum of the contributions from both
domains: 
\be  
A=\frac{2 I_1 \left(\omega \ln (\alpha's) 
\right)}{\omega \ln {(\alpha's)} }-1+
R \left(\alpha' s\right)^{\alpha(0)+\alpha't} ,
\label{pnp}
\ee
where $R$ is a constant and we have set $\mu^2=1/\alpha'$.

For infinite $s$ the first term on the right-hand side of \eq{pnp}
should dominate, so the Reggeon trajectory, which is linear for
$t>0$ with the slope $\alpha'\approx 1$~GeV from the measured
spectrum of mesons, has to be (almost) constant $\alpha(t)\approx 0.5$
for $t<0$, as it follows from the asymptotic behavior of $I_1$. 
Such a behavior would not agree with the experimental data for the
$\pi^0$ production in inclusive or exclusive processes at existing energies,
as was pointed out, respectively, in Refs.~\cite{BTT93,Kai06}.
At finite $s$ (but $s\gg -t$) both terms on the right-hand side of \eq{pnp} 
are important, while their relative strength depends on the 
value of $R$.

\begin{figure}
\vspace*{3mm}
\includegraphics[width=8cm]{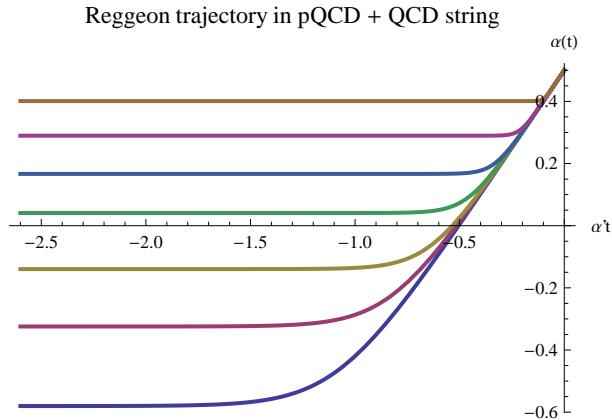} 
\caption[]{Effective $\rho$ Regge trajectory for $R= 23$.}   
\label{fi:plot}
\end{figure}

In Fig.~\ref{fi:plot} we
plot the effective Reggeon trajectory 
\be
\alpha_{\rm eff}(t)= \frac{\ln (A/R)}{\ln( \alpha' s)}
\ee
for various $\alpha's$:
$\alpha's =400$, $10^{3}$, $10^{4}$, $10^{6}$,  $10^{9}$, $10^{16}$, and $10^{40}$ 
from the bottom to the top.
The value of $R=23$ is taken to fit the data~\cite{Ken86} for 
$s=400~{\rm Gev}^2$, which are described by the lower line
in the figure. We have also substituted $\alpha(0)=0.5$ to agree
with the experimental data both for the meson masses and 
for the scattering processes.%
\footnote{This value of the Reggeon intercept is expected to emerge
due to spontaneous breaking of the chiral symmetry in QCD
and supersedes the one in \eq{alpha0}, though 
it is not yet shown how this happens.}

It is seen from Fig.~\ref{fi:plot} that for finite $s$ the linear Reggeon
trajectory continues from $t>0$ to $t<0$ and then flattens.
The smaller $s$ the deeper is the linear part at $t<0$.
A prediction of this figure is that $\alpha_{\rm eff}(t)\approx 0$
for $s=1~{\rm TeV}^2$. 
For $s$ larger than $10^3~{\rm TeV}^2$ it is rising with $s$ very slowly.

\section{Discussion and outlook}

We have considered in this Letter the two cases, 
where QCD string is outstretched:
the static potential and the Regge regime of meson scattering amplitudes.
They can be described by an effective string theory, which 
is an effective description for the degrees of freedom of long strings.
In this situation it does not matter what is the actual QCD string (if any)
which makes sense for all distances. 
Two such candidates for QCD string are:
\begin{itemize}
\addtolength{\itemsep}{-6pt}
\item  Migdal's elfin string~\cite{Mig81}, whose worldsheet is populated
by two-dimensional elementary fermions;
\item  holography%
\footnote{See Ref.~\cite{KS10}, where this approach is compared to the
lattice calculations.}
based on the AdS/CFT correspondence in a confining background, 
where extra degrees
of freedom are described by higher dimensions.
\end{itemize} 
In both cases the extra degrees of freedom are needed to provide
asymptotic freedom at small distances, but at the distances larger
than the confinement scale the mean-field approach apparently works
well, reproducing the same results as for the
Nambu--Goto string. This issue will be addressed elsewhere.

The ways the Dirichlet disk amplitude (reproducing the static potential)
and the scattering amplitude (reproducing the linear Reggeon trajectory)
are considered
are pretty much similar, although the latter refers, strictly speaking, to
scattering of tachyons, when the reparametrization path integral
decouples. In QCD we deal with off-shell scattering amplitudes,
when the  reparametrization path integral plays a crucial
role in maintaining the projective symmetry.
The path integral over reparametrizations obeying
$s(t_i)=s_i$ yields~\cite{MO10b}
\be
\frac{\int \D_{\rm diff} s \,G\left(s_j,s_j\right)}{\int \D_{\rm diff} s}
=\frac{1}{\pi}
\ln \frac{(s_{j+1}-s_{j-1})}{(s_{j+1}-s_{j})(s_{j}-s_{j-1})\varepsilon}
\ee
for the (singular) Green function at coinciding arguments, which 
is associated with the Lovelace Reggeon vertex
that results in consistent off-shell scattering amplitudes~\cite{DiV88},
reproducing the above results.

There is, however, a big difference in the square-root behavior
of the static potential~\rf{Vstring} and the linear Reggeon
trajectory~\rf{linear} of QCD string. The former is linked to the 
tachyon and the $d=1$ barrier, while the latter is apparently
associated with smooth surfaces. 

\begin{acknowledgments}
The subject of this Letter has been recently presented at the Seminars
at NBI, Nordita and ITEP. I am grateful to their participants for useful 
discussions.
\end{acknowledgments}


\end{document}